\documentclass[aps,pra,reprint,nofootinbib]{revtex4-1}

\usepackage{amsmath,amsfonts,amssymb,amsthm,bm,bbm,color}
\usepackage[colorlinks=true,linkcolor=blue,citecolor=magenta,urlcolor=blue]{hyperref}

\usepackage[pdftex]{graphicx}
\usepackage{subfig}

\usepackage[utf8]{inputenc}
\usepackage[T1]{fontenc}
\usepackage{lmodern}

\input{Qcircuit}

\renewcommand{\H}{\mathcal{H}}
\renewcommand{\1}{\mathbbm{1}}
\newcommand{\bra}[1]{\langle #1|}
\newcommand{\ket}[1]{|#1\rangle}
\newcommand{\braket}[2]{\langle #1 | #2 \rangle}
\newcommand{\su}{\mathfrak{su}}
\newcommand{\diag}{\operatorname{diag}}
\newcommand{\Tr}{\operatorname{tr}}

\begin{document}
\title{Self-testing properties of Gisin's elegant Bell inequality}
	
\author{Ole Andersson}
	\email{ole.andersson@fysik.su.se}
\author{Piotr Badzi\c{a}g}
	\email{piotr.badziag@gmail.com}
\author{Ingemar Bengtsson}
	\email{ingemar.bengtsson@fysik.su.se}
\author{Irina Dumitru}
	\email{irina.dumitru@fysik.su.se}
\affiliation{Fysikum, Stockholms Universitet, 106 91 Stockholm, Sweden}
\author{Ad\'{a}n Cabello}
	\email{adan@us.es}
\affiliation{Departamento de F\'{i}sica Aplicada II, Universidad de Sevilla, 41012 Sevilla, Spain}
	
\begin{abstract}
An experiment in which the Clauser-Horne-Shimony-Holt inequality is maximally violated is self-testing (i.e., it certifies in a device-independent way both the state and the measurements).
We prove that an experiment maximally violating Gisin's elegant Bell inequality is not similarly self-testing.
The reason can be traced back to the problem of distinguishing an operator from its complex conjugate.
We provide a complete and explicit characterization of all scenarios in which 
the elegant Bell inequality is maximally violated. This enables us to see exactly how the problem plays out.
\end{abstract}

\date{\today}

\maketitle

\section{Introduction}

Bell inequalities are correlation inequalities which are satisfied by any local realistic model
but can be violated by quantum theory \cite{Bell64}. They thus allow us to test the former 
against the latter. They are also useful in practical applications like secure communication \cite{Ekert91}, reduction of 
communication complexity \cite{BZPZ04}, and secure private randomness \cite{Colbeck06}. 
For such applications, the self-testing properties of some Bell inequalities play a major role,
as they allow a maximal quantum violation to occur in an effectively unique way.
In the current paper we investigate the self-testing properties implied by a maximal violation of the so-called
elegant Bell inequality (EBI). 

\begin{figure}[t]
	\centering
	\subfloat[The octahedron in \newline Alice's Bloch sphere.\label{octahedron}]{%
		\includegraphics[width=0.22\textwidth]{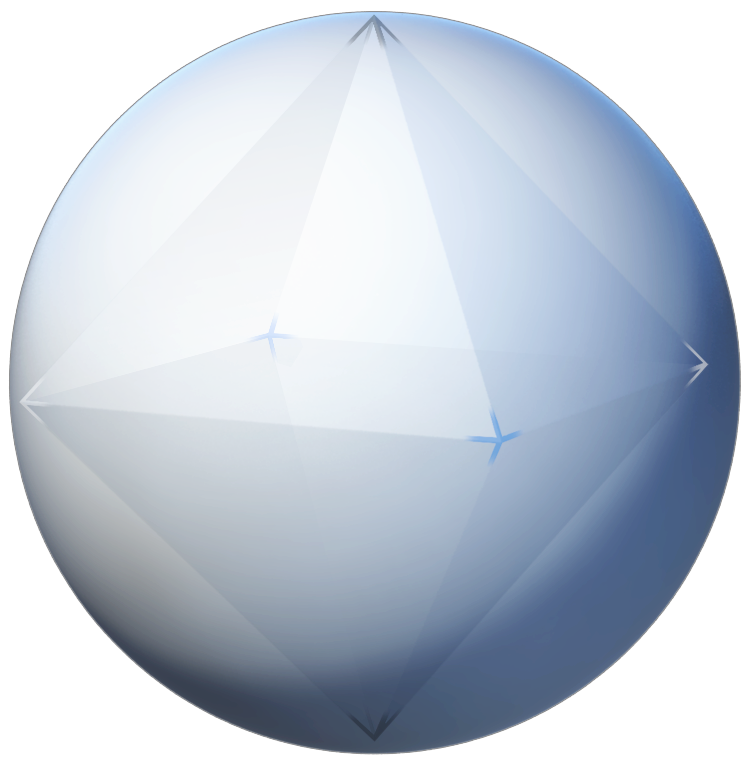}
	}
	~\hspace{5pt}
	\subfloat[The cube in Bob's \newline Bloch sphere.\label{cube}]{%
		\includegraphics[width=0.22\textwidth]{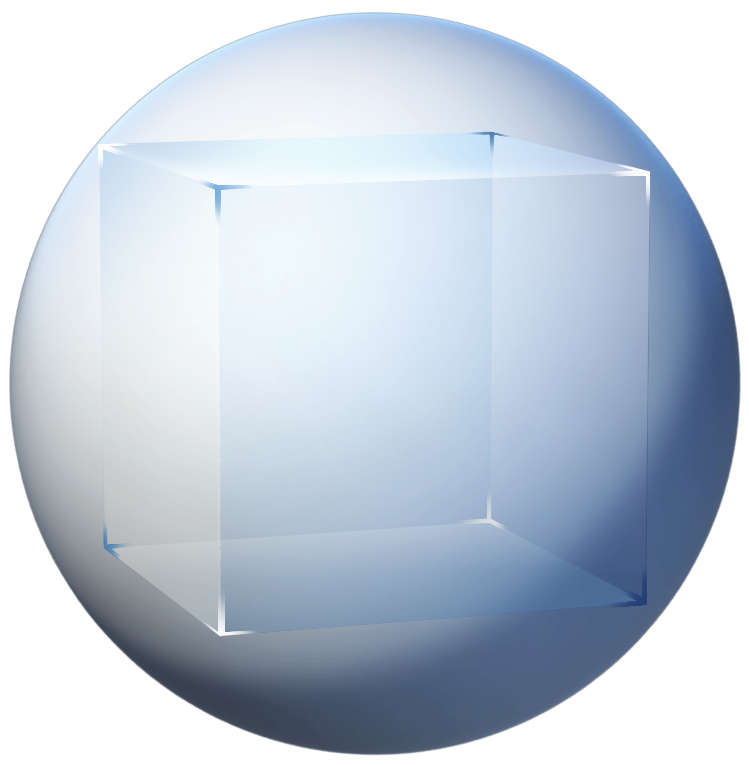}
	}
\caption{Alice's and Bob's measurement eigenstates form two dual Platonic solids inscribed in Bloch spheres. 
Alice's eigenstates sit at the corners of an octahedron, 
Bob's eigenstates can be grouped into two dual sets of SIC vectors which sit at the corners of a cube.} 
	\label{Figure}
\end{figure}

The EBI involves two parties, Alice and Bob, measuring three and four dichotomic observables, respectively.
If the possible outcomes of these observables are taken to be $-1$ and $+1$, and we write $E_{kl}$ for the 
expectation value of the product of the outcomes of Alice's $k$th observable and Bob's $l$th observable,
the EBI reads 
\begin{equation}\label{EBI}
	\begin{split}
		S \equiv E_{11}&+E_{12}-E_{13}-E_{14}+E_{21}-E_{22}\\
		               &+E_{23}-E_{24}+E_{31}-E_{32}-E_{33}+E_{34}\leq 6.
	\end{split}
\end{equation}
The EBI does not define a facet of the classical correlation polytope 
and, therefore, it does not reflect the geometry of the latter. Rather, according to 
Gisin \cite{Gisin2009}, its elegance resides in the way it is maximally violated by 
quantum theory. The maximum violation, proven to be $S=4\sqrt{3}$ by Ac\'{\i}n {\em et al.}\ \cite{Acin2016}, occurs when Alice and Bob use 
projective measurements whose eigenstates are maximally 
spread out on Bloch spheres, in a sense made precise below. In the particular case when they share 
a two-qubit state,
Alice's measurement eigenstates form a complete set 
of three mutually unbiased bases (MUBs), while those of Bob are eight states that can be 
partitioned into two dual sets of SIC elements,
see Fig.\ \ref{Figure}. SICs are also known as symmetric informationally complete positive operator-valued measures (SIC-POVMs).
However, here the configuration arises from four projective measurements and not from two POVMs.
Since MUBs (and SICs)
are intriguing configurations of independent interest \cite{Wootters2006}, we can 
ask the question:
does maximum quantum violation of the EBI \emph{require} 
the existence of three MUBs in dimension two, with no assumptions about the 
preparation and measurement devices being made? 

There is another motivation of more immediate practical relevance. 
Recently, Ac\'{\i}n {\em et al.}\ \cite{Acin2016} addressed the problem of how to 
use a two-qubit entangled state together with a local POVM measurement to certify 
the generation of two bits of device-independent private randomness. 
They provided two methods for such a certification.
The simplest one was based on the EBI, and was supported by numerical results.
They suggested that an analytical proof of the correctness of the method should 
rely on a proof that a maximal violation of the EBI self-tests the maximally 
entangled state and the three Pauli measurements that give rise to the MUB. 

In this paper we will prove that the EBI does {\em not} provide a self-test for the maximally entangled state and the three Pauli measurements, in the strict sense of  
Refs.\ \cite{McKague2010, McKague2010thesis}. It comes close to doing so though and we  
discuss the implications for the method suggested by Ac\'{\i}n {\em et al.}\ in a separate paper \cite{Andersson2017}.
In Sec.\ \ref{Self} of this paper we review the strict definition of self-testing. 
In Sec.\ \ref{When} we discuss, following Refs.\ \cite{Acin2016, Popescu1992}, maximal violation of the EBI. 
Section \ref{SelfEBI} contains our main results on the self-testing properties of the EBI. To make the 
paper easier to read some of the detailed derivations are given in Sec.\ \ref{Proofs}. Finally, 
Sec.\ \ref{Conclusion} states our conclusions and the outlook.
 
\section{Self-testing experiments}\label{Self}
The concept of \emph{self-testing} was introduced by Mayers and Yao \cite{Mayers1998} 
as a test for a photon source which, if passed, guarantees that the source is 
adequate for the security of the BB84 protocol for quantum key distribution. 
Self-testing then received a stringent definition by the same authors in Ref.\ \cite{Mayers2004}, 
a definition which was further polished by Magniez {\em et al.}\ \cite{Magniez2006} 
and McKague and Mosca \cite{McKague2010, McKague2010thesis}. In this paper we 
adopt the definition of self-testing used in these latter references.

The definition of being self-testing consists of a condensed description of how 
a \emph{reference experiment} can be modified without affecting the statistics. 
Allowed modifications include local rotations, addition of ancillas, changes of 
the effect of observables outside the support of the state, and local embeddings 
of states and observables into greater or smaller Hilbert spaces \cite{McKague2010,McKague2010thesis}.
Here we give the definition at a level of generality sufficient for our purposes. 
We thus consider a reference experiment involving two parties, Alice and Bob, 
performing $m$ and $n$ local dichotomic measurements $a_k=\{\Pi_{\pm}^{a_k}\}$ 
and $b_l=\{\Pi_{\pm}^{b_l}\}$, respectively, on a given bipartite state $\ket{\phi}$.
(The subscript signs label the measurement outcomes.)
We then say that the reference experiment is \emph{self-testing} if for any other 
experiment in which Alice performs $m$ local measurements $A_k=\{\Pi_{\pm}^{A_k}\}$ 
and Bob performs $n$ local measurements $B_l=\{\Pi_{\pm}^{B_l}\}$
on a shared state $\ket{\psi}$, a complete agreement of the two experiments statistics, i.e., equality 
\begin{equation}
	\bra{\phi} \Pi_{\pm}^{a_k} \Pi_{\pm}^{b_l} \ket{\phi} = \bra{\psi} \Pi_{\pm}^{A_k} \Pi_{\pm}^{B_l} \ket{\psi}
\end{equation}
for all $k,l$, implies the existence of a local unitary, or, more precisely, a local isometric embedding
\begin{equation}
	\begin{split}
		\Phi=\Phi_A\otimes \Phi_B : \H_A \otimes \H_B \to &\,(\H_A\otimes \H_a) \otimes (\H_B\otimes\H_b) \\
		                                                = &\, (\H_A\otimes \H_B)\otimes(\H_a\otimes\H_b)
	\end{split}	                                                
\end{equation}
such that $\Phi(\Pi_{\pm}^{A_k}\Pi_{\pm}^{B_l}\ket{\psi})=\ket{\chi}\otimes \Pi_{\pm}^{a_k}\Pi_{\pm}^{b_l}\ket{\phi}$,
where $\ket{\chi}$ is some arbitrary but normalized `junk' vector in $\H_A\otimes \H_B$.
(Here we use vocabulary introduced in Refs.\ \cite{McKague2010,McKague2010thesis}.)
Notice that the definition of self-testing captures, although in a rather abstract way, 
the physical intuition that the state generation includes a successful isolation of a `relevant part'
of the total state. On this part, the measurements then act in a way stipulated by the reference experiment 
without entangling it with the rest of the state.
We emphasize this by saying, for short, that the experiment is \emph{effectively equivalent} to the reference experiment.

\section{Maximal violation of the EBI}\label{When}
The elegant Bell inequality can be violated in quantum theory.
In fact, Ac\'{\i}n {\em et al.}\ \cite{Acin2016} have recently proven that the maximum quantum value that $S$ can attain is $4\sqrt{3}$.
The simplest setting when this happens, it turns out, is when Alice and Bob share two qubits in the maximally entangled state   
\begin{equation}\label{singlet}
\ket{\phi_+}=\tfrac{1}{\sqrt{2}}(\ket{0_a0_b}+\ket{1_a1_b}), 
\end{equation}
Alice's observables correspond to the three Pauli operators
\begin{equation}\label{a}
	a_1=Z=\sigma_Z, \qquad a_2=X=\sigma_X, \qquad a_3=Y=\sigma_Y,
\end{equation}
and Bob's observables correspond to 
\begin{subequations}\label{b}
\begin{alignat}{2}
b_1 &= \tfrac{1}{\sqrt{3}}(Z+X-Y),\qquad & b_3 &= \tfrac{1}{\sqrt{3}}(-Z+X+Y), \label{b1} \\
b_2 &= \tfrac{1}{\sqrt{3}}(Z-X+Y),\qquad & b_4 &= \tfrac{1}{\sqrt{3}}(-Z-X-Y). \label{b2} 
\end{alignat}
\end{subequations}

The elegance of the Bell inequality \eqref{EBI} is apparent \cite{Gisin2009} when we observe 
that the observables in Eqs.\ \eqref{a} and \eqref{b} give rise to two measurement 
structures which can be represented by two dual polyhedra in the Bloch ball:
Alice's measurement eigenstates form a complete set 
of three MUBs, with each basis corresponding to a pair of opposite corners
of an octahedron inscribed in the Bloch sphere, see Fig.\ \ref{octahedron}. On the Bloch sphere,
the eight eigenstates of Bob's projective measurements   
form the vertices of a dual cube, see Fig.\ \ref{cube}. They can be grouped into two tetrahedra containing no adjacent corners. The vertices of such a tetrahedron
can be regarded as the four vectors in a SIC, and we can arrange them such that one SIC is formed by the $-1$ outcome projectors and the other by the $+1$ outcome projectors. 
Below we will show that, in general, the EBI is maximally violated if, and only if, the state is a superposition of
maximally entangled qubit states like the one in Eq.\ \eqref{singlet} and Alice's and Bob's observables split 
into direct sums of qubit MUB-SIC configurations similar to that just described.  

To characterize all scenarios in which the EBI is maximally violated we consider a general one in
which Alice measures three dichotomic observables $A_1,A_2,A_3$ 
and Bob measures four dichotomic observables $B_1,B_2,B_3,B_4$, all of which take the 
values $-1$ or $+1$, on a bipartite system in a state $\ket{\psi}$ such that 
$\bra{\psi}\Sigma\ket{\psi}=4\sqrt{3}$, where $\Sigma$ is the \emph{elegant Bell operator}:
\begin{equation}\label{EBO}
	\begin{split}
		\hspace{-3pt}\Sigma 
		\equiv& A_1 B_1+A_1 B_2-A_1 B_3-A_1 B_4+A_2 B_1-A_2 B_2 \\
		 &+A_2 B_3-A_2 B_4+A_3 B_1-A_3 B_2-A_3 B_3+A_3 B_4.
	\end{split}
\end{equation}

The first assertion, which, like all other assertions in this section, is proven in Sec.\ \ref{Proofs}, is that
Alice's and Bob's observables preserve the supports, even the eigenspaces, of the respective marginal states: If $\lambda_1,\lambda_2,\dots,\lambda_m$ are the \emph{different} Schmidt coefficients 
of $\ket{\psi}$, having multiplicities $d_1,d_2,\dots, d_m$, and $\H_A^i$ and $\H_B^i$ denote 
the $d_i$-dimensional eigenspaces of $\Tr_B\ket{\psi}\bra{\psi}$ and $\Tr_A\ket{\psi}\bra{\psi}$ 
corresponding to the eigenvalue $\lambda_i^2$, then Alice's observables 
send $\H_A^i$ into itself and Bob's observables send $\H_B^i$ into itself. As a consequence we can, without loss of generality, truncate Alice's and Bob's Hilbert spaces and restrict the observables to the support of the respective marginal state. 
We henceforth assume this has been done and we write 
$A_k^i$ and $B_l^i$ for the restriction of Alice's $k$th and Bob's $l$th observable to $\H_A^i$ and 
$\H_B^i$, respectively.

The second assertion is that 
Alice's observables anti-commute: $\{A_k,A_l\}=2\delta_{kl}$. 
(Since their eigenvalues equal $-1$ or $+1$, Alice's and Bob's observables are 
involutions, i.e., they square to the identity operator.) 
From this follows that $\H_A^i$ 
is even-dimensional, say $d_i=2n_i$, and can be split into $2$-dimensional and 
pairwise orthogonal subspaces, each left invariant by Alice's observables:
\begin{equation}\label{sitA}
	\H_A^i = \bigoplus_{p=1}^{n_i}\H_A^{ip}, \qquad 
	A_k^i = \bigoplus_{p=1}^{n_i} A_k^{ip}.
\end{equation}
Furthermore, each subspace $\H_A^{ip}$ admits a basis $\{\ket{0_A^{ip}},\ket{1_A^{ip}}\}$
with respect to which
\begin{equation}\label{exactA}
	A_1^{ip} = Z, \qquad A_2^{ip} = X, \qquad A_3^{ip} = \pm Y.
\end{equation}
Notice the indefinite sign of $A_3^{ip}$; a similar sign indeterminacy was identified in 
\cite{McKague2010}, treating a related problem.

The third assertion is that every $\H_B^i$ can as well be decomposed into $2$-dimensional orthogonal subspaces, each of which is left invariant 
by Bob's observables:
\begin{equation}
	\H_B^i = \bigoplus_{p=1}^{n_i} \H_B^{ip}, \qquad 
	B_l^i = \bigoplus_{p=1}^{n_i} B_l^{ip}.
\end{equation}
Moreover, $\H_B^{ip}$ admits a basis $\{\ket{0_B^{ip}},\ket{1_B^{ip}}\}$ such that,
as matrices with respect to $\{\ket{0_A^{ip}},\ket{1_A^{ip}}\}$ and $\{\ket{0_B^{ip}},\ket{1_B^{ip}}\}$,
\begin{subequations}\label{B}
\begin{align}
	B_1^{ip} &= \tfrac{1}{\sqrt{3}}(A_1^{ip} + A_2^{ip} - A_3^{ip})   = \tfrac{1}{\sqrt{3}}(Z + X \mp Y),   \label{B1} \\
	B_2^{ip} &= \tfrac{1}{\sqrt{3}}(A_1^{ip} - A_2^{ip} + A_3^{ip})   = \tfrac{1}{\sqrt{3}}(Z - X \pm Y),   \label{B2} \\
	B_3^{ip} &= \tfrac{1}{\sqrt{3}}(- A_1^{ip} + A_2^{ip} + A_3^{ip}) = \tfrac{1}{\sqrt{3}}(- Z + X \pm Y), \label{B3} \\
	B_4^{ip} &= \tfrac{1}{\sqrt{3}}(- A_1^{ip} - A_2^{ip} - A_3^{ip}) = \tfrac{1}{\sqrt{3}}(- Z - X \mp Y). \label{B4}
\end{align}
\end{subequations}

The fourth and last assertion concerns the state.
The bases $\{\ket{0_A^{ip}},\ket{1_A^{ip}}\}$ and 
$\{\ket{0_B^{ip}},\ket{1_B^{ip}}\}$ are eigenbases of Alice's and Bob's local states 
which will be constructed in such a way that the shared state obtains the representation
\begin{equation}\label{state}
\begin{split}
	\ket{\psi} 
	&= \sum_{i=1}^{m}\sum_{p=1}^{n_i}\lambda_i (\ket{0_A^{ip} 0_B^{ip}} + \ket{1_A^{ip} 1_B^{ip}})\\
	&= \sqrt{2}\sum_{i=1}^{m}\sum_{p=1}^{n_i}\lambda_i\ket{\phi_+^{ip}}.
\end{split}
\end{equation}
Notice that $\ket{\phi_+^{ip}}$ is the Einstein-Podolsky-Rosen singlet in the space $\H_A^{ip}\otimes \H_B^{ip}$, restricted to which
Alice's and Bob's observables are given by Eqs.\ \eqref{exactA} and \eqref{B}.
For each $i$, we arrange that $A_3^{ip}=Y$ for $p\leq r_i$ and $A_3^{ip}=-Y$ for $p>r_i$, where $0\leq r_i\leq n_i$.
For any Schmidt coefficients $\lambda_i$ and any $r_i$ the EBI is maximally violated.

We end this section with some remarks about mixed states and general measurements
violating the EBI maximally.
If Alice and Bob share a mixed state which can be expanded as an incoherent sum of pure states, each of which individually maximally violates the EBI, then so does the mixed state. A straightforward convexity argument then shows that this is the only possibility for a mixed state violating the EBI maximally. 
One can also ask if the EBI can be maximally violated by nonprojective measurements.
It turns out that this is not possible. More precisely, if Alice and Bob measures local dichotomic POVMs and the EBI is maximally violated, then the measurement operators preserve the supports of the local states, and when restricted to these supports the measurements are projective. A proof of this can be based on Naimark's dilation theorem (see, e.g., \cite{Holevo2011}) and the arguments in the second paragraph in Sec.\ \ref{Proofs} below.

\section{Self-testing properties of the EBI}\label{SelfEBI}
By the previous section, Alice's observables split into an unknown number of $2$-dimensional $\su(2)$ representations 
and an unknown number of `transposed' $\su(2)$ representations. The statistics, however, is independent of these numbers, since 
the statistics equals that of the experiment specified by Eqs.\ \eqref{singlet}-\eqref{b}, from now on referred to 
as `the reference experiment'. The reference experiment is therefore not self-testing, and neither is 
any other experiment in which only a maximal violation of the EBI is assumed.
For if a local isometric embedding $\Phi$ exists, establishing an effective equivalence between the reference 
experiment and the generic experiment in Sec.\ \ref{When}, then
\begin{equation}\label{necessary}
	\begin{split}
		\bra{\phi_+}a_2 a_3 &(b_1 +b_2)\ket{\phi_+} = \\
		                         &= \braket{\Phi(A_2\ket{\psi})}{\Phi(A_3(B_1+B_2)\ket{\psi})} \\
											&= \bra{\psi}A_2A_3(B_1+B_2)\ket{\psi}.
	\end{split}
\end{equation}
But $\bra{\phi_+}a_2a_3(b_1+b_2)\ket{\phi_+}=2i/\sqrt{3}$
and
\begin{equation}
	\bra{\psi}A_2A_3(B_1+B_2)\ket{\psi}=\frac{2i}{\sqrt{3}}\sum_{i=1}^m \lambda_i^2(4r_i-2n_i).\label{stora}
\end{equation}
The results agree if and only if $r_i = n_i$ for all $i$. (Remember that $2n_i$ is the multiplicity of the Schmidt coefficient $\lambda_i$.) But, because the values of the differences $n_i-r_i$ are not 
determinable from the statistics of the experiment, this shows that 
a maximal violation of the EBI is not sufficient to conclude that the reference experiment is self-testing.

On the other hand, if we \emph{require} that Eq.\ \eqref{necessary} \emph{is} satisfied, in addition to a maximal violation of the EBI, the reference experiment \emph{is} self-testing; an equivalence is provided by the local isometric embedding $\Phi$ given by the circuit 
\begin{equation*}
\hspace{30pt}
\Qcircuit @C=0.4cm @R=0.3cm {
	\lstick{\ket{0_a}}                        & \gate{H}   & \ctrl{1}    & \gate{H}   & \ctrl{1}    & \qw \\
	&                                       & \gate{A_1} & \qw         & \gate{A_2} & \qw         &     \\
	\lstick{\ket{\psi}}\qwx[1,1]\qwx[-1,1]  &            &             &            &             &     \\
	&                                       & \gate{\frac{\sqrt{3}}{2}(B_1+B_2)} & \qw         & \gate{\frac{\sqrt{3}}{2}(B_1+B_3)} & \qw         &     \\
	\lstick{\ket{0_b}}                        & \gate{H}   & \ctrl{-1}   & \gate{H}   & \ctrl{-1}   & \qw \\
}
\end{equation*}
(Here $H$ denotes the Hadamard gate and the control gates are triggered by the presence of $\ket{1_a}$ and $\ket{1_b}$.) 
McKague and Mosca used this isometric embedding to develop a generalized Mayers-Yao test, see \cite{McKague2010},
and McKague {\em et al.}\ \cite{McKague2012} used it to show that the standard scenario in which the Clauser-Horne-Shimony-Holt (CHSH) Bell 
inequality is maximally violated is robustly self-testing. Recently, a more universal form of this isometric embedding 
was used to prove that all pure bipartite entangled states can be self-tested \cite{Coladangelo2016}.

Straightforward calculations show that
\begin{equation}
	\Phi \big( \Pi^{A_k}_{\pm} \Pi^{B_l}_{\pm} \ket{\phi_+^{ip}} \big)
	=\ket{0^{ip}_A0^{ip}_B}\otimes \Pi^{a_k}_{\pm}\Pi^{b_l}_{\pm}
			\ket{\phi_+},
\end{equation}
where $\Pi^{A_k}_{\pm}$ and $\Pi^{B_l}_{\pm}$ are the projections onto the $\pm 1$-eigenspaces of $A_k$ and $B_l$, 
and $\Pi^{a_k}_{\pm}$ and $\Pi^{b_l}_{\pm}$ are the projections onto the $\pm 1$-eigenspaces of 
the observables $a_k$ and $b_l$ in the reference experiment. Consequently,
\begin{equation}\label{calc}
	\begin{split}
		\Phi \big( \Pi^{A_k}_{\pm} \Pi^{B_l}_{\pm} \ket{\psi} \big) 
		&= \sqrt{2} \sum_{i=1}^m \sum_{p=1}^{n_i} \lambda_i \ket{0^{ip}_A0^{ip}_B} \otimes \Pi^{a_k}_{\pm}\Pi^{b_l}_{\pm}
			\ket{\phi_+} \\
		&= \ket{\chi} \otimes \Pi^{a_k}_{\pm}\Pi^{b_l}_{\pm}\ket{\phi_+}.
	\end{split}
\end{equation}

The last identity in Eq.\ \eqref{calc} defines the junk vector $\ket{\chi}$.
If Eq.\ \eqref{necessary} is \emph{not} satisfied, the junk vector naturally splits into two parts, 
$\ket{\chi}=\ket{\chi_1}+\ket{\chi_2}$, defined by
\begin{align}
	\ket{\chi_1} &= \sqrt{2} \sum_{i=1}^m \sum_{p=1}^{r_i} \lambda_{i} \ket{0^{ip}_A0^{ip}_B}, \\
	\ket{\chi_2} &= \sqrt{2} \sum_{i=1}^m \sum_{p=r_i+1}^{n_i} \lambda_{i} \ket{0^{ip}_A0^{ip}_B}.
\end{align}
Equation \eqref{calc} is then no longer valid. Instead we have that 
\begin{subequations}
\begin{align}
	&\hspace{-10pt}\Phi \big( \Pi^{A_1}_{\pm} \Pi^{B_l}_{\pm} \ket{\psi} \big) 
		= \ket{\chi_1} \Pi_{\pm}^{a_1} \Pi_{\pm}^{b_l} \ket{\phi_+} 
		+ \ket{\chi_2} \Pi_{\pm}^{a_1} \Pi_{\mp}^{b_{5-l}} \ket{\phi_+}, \\
	&\hspace{-10pt}\Phi \big(\Pi^{A_2}_{\pm} \Pi^{B_l}_{\pm} \ket{\psi} \big) 
		= \ket{\chi_1} \Pi_{\pm}^{a_2} \Pi_{\pm}^{b_l} \ket{\phi_+} 
		+ \ket{\chi_2} \Pi_{\pm}^{a_2} \Pi_{\mp}^{b_{5-l}} \ket{\phi_+}, \\
	&\hspace{-10pt}\Phi \big( \Pi^{A_3}_{\pm} \Pi^{B_l}_{\pm} \ket{\psi} \big) 
		= \ket{\chi_1} \Pi_{\pm}^{a_3} \Pi_{\pm}^{b_l} \ket{\phi_+} 
		+ \ket{\chi_2} \Pi_{\mp}^{a_3} \Pi_{\mp}^{b_{5-l}} \ket{\phi_+}.
\end{align}
\end{subequations}
Using these identities one can show that a measurement of Alice's third observable,
 or a measurement of any of Bob's observables, entangles  
the singlet part of the state with the junk part. But, interestingly, even though an adversary, Eve, 
having access only to the junk part, can detect a measurement of $A_3$ or any of the $B_l$s, she 
cannot distinguish between the outcomes. This is so because, irrespective of the measurement outcome, all these measurements leave Eve's system 
in the same state. 

\section{Derivations}\label{Proofs}
In this section we prove the assertions in 
Sec.\ \ref{When}. Inspiration comes mainly from 
Ac\'{\i}n {\em et al.}'s derivation of the least quantum bound for the EBI \cite{Acin2016} 
and from Popescu and Rohrlich's characterization of the scenarios 
in which the CHSH Bell inequality is maximally violated \cite{Popescu1992}.

First we prove that Alice's and Bob's observables preserve the supports of the marginal states. Thus let $\ket{\psi}$ be a state saturating the EBI and let 
$\ket{\psi}=\sum_{i=1}^{m} \sum_{p=1}^{d_i}\lambda_i\ket{u_p^iv_p^i}$
be a Schmidt decomposition, 
with $i$ labeling the $m$ different Schmidt coefficients and $d_i$ being the multiplicity 
of $\lambda_i$. Define 
\begin{subequations}\label{D}
\begin{align}
	D_1 &= \tfrac{1}{\sqrt{3}}(A_1 + A_2 + A_3), \label{D1} \\
	D_2 &= \tfrac{1}{\sqrt{3}}(A_1 - A_2 - A_3), \label{D2} \\
	D_3 &= \tfrac{1}{\sqrt{3}}(-A_1+A_2-A_3),    \label{D3} \\
	D_4 &= \tfrac{1}{\sqrt{3}}(-A_1-A_2+A_3).    \label{D4}
\end{align}
\end{subequations}
Then $\sum_{l=1}^4 (D_l-B_l)^2=8\1-2\Sigma/\sqrt{3}$ and, hence, 
\begin{equation}\label{kill}
	\sum_{i=1}^m \sum_{p=1}^{d_i} \lambda_i D_l\ket{u^i_pv^i_p}
	= \sum_{i=1}^m \sum_{p=1}^{d_i} \lambda_i B_l\ket{u^i_pv^i_p}.
\end{equation}
Multiplication of both sides by $\bra{w,v^{j}_{q}}$, 
where $\ket{w}$ is any vector in $\H_A$ perpendicular to the support of $\Tr_B\ket{\psi}\bra{\psi}$,
yields the identity $\lambda_{j} \bra{w}D_l\ket{u^{j}_{q}}=0$. Since the indices $j$ 
and $q$ are arbitrary and $\lambda_{j}>0$, this proves that 
$D_l$ preserves the support of $\Tr_B\ket{\psi}\bra{\psi}$.
Then so does each $A_k$. A similar argument shows that the 
operators $B_l$ preserve the support of 
the marginal state $\Tr_A\ket{\psi}\bra{\psi}$.

Next we prove that Alice's and Bob's observables preserve the 
eigenspaces of the marginal states. 
From Eq.\ \eqref{kill} follows that for any two pairs of indices $(i_1,p_1)$ and $(i_2,p_2)$,
\begin{equation}
	\lambda_{i_2}  \bra{u_{p_1}^{i_1}}D_l\ket{u_{p_2}^{i_2}}
	=\lambda_{i_1} \bra{v_{p_2}^{i_2}}B_l\ket{v_{p_1}^{i_1}}. \label{transpose}
\end{equation}
This, in turn, implies that
\begin{equation} 
	\lambda_{i_1}^2  \bra{u_{p_1}^{i_1}}D_l\ket{u_{p_2}^{i_2}}
	=\lambda_{i_2}^2 \bra{u_{p_1}^{i_1}}D_l\ket{u_{p_2}^{i_2}}. \label{block}
\end{equation}
From Eq.\ \eqref{block} we can deduce that $D_l$ and, hence, each $A_k$ preserves the eigenspaces $\H_A^i$.
By an identical argument also the 
operators $B_l$ preserve the eigenspaces $\H_B^i$.
We write $A_k^i$ and $D_l^i$ for the restrictions of $A_k$ and $D_l$ to $\H_A^i$,
and $B_l^i$ for the restriction of $B_l$ to $\H_B^i$.

From Eq.\ \eqref{D} and the $A_k$s being involutions follow that
\begin{subequations}\label{DD}
\begin{align}
	(D_1^i)^2 &= \1 + \tfrac{1}{3}(\{A_1^i,A_2^i\}+\{A_1^i,A_3^i\}+\{A_2^i,A_3^i\}), \label{DD1}\\
	(D_2^i)^2 &= \1 - \tfrac{1}{3}(\{A_1^i,A_2^i\}-\{A_1^i,A_3^i\}+\{A_2^i,A_3^i\}), \label{DD2}\\
	(D_3^i)^2 &= \1 - \tfrac{1}{3}(\{A_1^i,A_2^i\}+\{A_1^i,A_3^i\}-\{A_2^i,A_3^i\}), \label{DD3}\\
	(D_4^i)^2 &= \1 + \tfrac{1}{3}(\{A_1^i,A_2^i\}-\{A_1^i,A_3^i\}-\{A_2^i,A_3^i\}). \label{DD4}
\end{align}
\end{subequations}
Furthermore, from Eq.\ \eqref{transpose} and each $B_l$ being an involution follows that
$D_l^i$ is an involution. But then, by Eq.\ \eqref{DD},
\begin{equation}\label{anticommutation}
	\{A_1^i,A_2^i\}=\{A_1^i,A_3^i\}=\{A_2^i,A_3^i\}=0.
\end{equation}

Equation \eqref{anticommutation} implies that $A_1^i$, $A_2^i$, and $[A_1^i,A_2^i]/2i$ generate 
an $\su(2)$ representation. We cannot, however, conclude that 
$A_3^i=[A_1^i,A_2^i]/2i$. Nevertheless, among the irreducible $\su(2)$ representations only the 
$2$-dimensional one satisfies Eq.\ \eqref{anticommutation}. The space $\H_A^i$ must therefore 
be even-dimensional, say $d_i=2n_i$, and be decomposable into an orthogonal direct sum of 
$2$-dimensional subspaces, $\H_A^i=\bigoplus_{p=1}^{n_i}\H_A^{ip}$,
each of which is left invariant by $A_1^i$ and $A_2^i$; thus $A_1^i = \bigoplus_{p=1}^{n_i}A_1^{ip}$ and $A_2^i=\bigoplus_{p=1}^{n_i}A_2^{ip}$.
Furthermore, since $A_1^i$ and $A_2^i$ are involutions, we can choose a provisional basis 
$\{\ket{s_A^i}\}_{s=1}^{d_i}$ in each $\H_A^i$ such that for every $1\leq p\leq n_i$, $\{\ket{(2p-1)_A^{i}},\ket{(2p)_A^{i}}\}$ is a basis in $\H_A^{ip}$ relative to which $A_1^{ip}=Z$ and $A_2^{ip}=X$.

It remains to prove that the decomposition of $\H_A^i$ can be chosen such that $A_3^i$ 
also splits into a direct sum, $A_3^i=\bigoplus_{p=1}^{n_i}A_3^{ip}$,
and that the basis in $\H_A^{ip}$ can be chosen such that $A_3^{ip}=\pm Y$. 
To this end, let $(A_3^i)_{p_2}^{p_1}$ be the $2\times 2$ matrix which in the provisional basis describes 
how $A_3^i$ connects $\H_A^{ip_1}$ to $\H_A^{ip_2}$. Then, by Eq.\ \eqref{anticommutation}, 
and since $A_3^i$ is Hermitian, $(A_3^i)_{p_2}^{p_1}=\omega_{p_2}^{p_1}Y$ for some real number $\omega_{p_2}^{p_1}$.
Next introduce a tensor product structure in $\H_A^i$ by writing 
$\ket{(2p-1)_A^{i}}=\ket{p}\otimes\ket{0}$ and $\ket{(2p)_A^{i}}=\ket{p}\otimes\ket{1}$. Then
$A_1^i=\1\otimes Z$, $A_2^i=\1\otimes X$, and $A_3^i=\Omega\otimes Y$, 
where $\Omega$ is the $n_i\times n_i$ matrix whose element on position $(p_1,p_2)$ is $\omega_{p_2}^{p_1}$.
Being Hermitian, $\Omega$ can be diagonalized, say $U^\dagger\Omega U=\diag(\omega_1,\omega_2,\dots,\omega_{n_i})$. Then
\begin{subequations}
\begin{align}
	(U^\dagger\otimes\1) A_1^i (U\otimes\1) &= \1\otimes Z,\\ 
	(U^\dagger\otimes\1) A_2^i (U\otimes\1) &= \1\otimes X,\\
	(U^\dagger\otimes\1) A_3^i (U\otimes\1) &= \diag(\omega_1,\omega_2,\dots,\omega_{n_i})\otimes Y.
\end{align}
\end{subequations}
Each diagonal element $\omega_p$ equals $+1$ or $-1$ because $A_3^i$ is an involution.
We choose $U$ such that $\omega_p=+1$ for $p\leq r_i$ and $\omega_p=-1$ for $p>r_i$, where $r_i$ is the number of positive diagonal elements.
We then rotate the provisional basis by applying $U^\dagger\otimes \1$ to it
and rotate the $\H_A^{ip}$s accordingly.
 
Next we consider Bob's observables. These are completely determined by Alice's observables.
To see this, define 
\begin{equation}
	\ket{s_B^i} = \sum_{p=1}^{n_i} \ket{v_p^i} \braket{s_A^i}{u_p^i}.
\end{equation}
Then $\bra{s_B^i}B_l^i\ket{t_B^{i}}=\bra{t_A^{i}}D_l^i\ket{s_A^{i}}$ and, hence, by Eq.\ \eqref{D},
\begin{subequations}
\begin{align}
	B_1^i &= \tfrac{1}{\sqrt{3}}(A_1^i+A_2^i+A_3^i)^T,  \\
	B_2^i &= \tfrac{1}{\sqrt{3}}(A_1^i-A_2^i-A_3^i)^T,  \\
	B_3^i &= \tfrac{1}{\sqrt{3}}(-A_1^i+A_2^i-A_3^i)^T, \\
	B_4^i &= \tfrac{1}{\sqrt{3}}(-A_1^i-A_2^i+A_3^i)^T. 
\end{align}
\end{subequations}
This proves Eq.\ \eqref{B}.

The assertion about the state is a straightforward consequence of the calculation
\begin{equation}
	\begin{split}
		\hspace{-5pt}\ket{\psi}
		&= \sum_{i=1}^{m} \sum_{p=1}^{d_i} \lambda_i \ket{u_p^iv_p^i} \\
		&= \sum_{i=1}^{m} \sum_{p=1}^{d_i} \sum_{s=1}^{d_i} \sum_{t=1}^{d_i} 
		   \lambda_i \ket{s_A^i t_B^i} \braket{s_A^i}{u_p^i} \braket{t_B^i}{v_p^i} \\
		&= \sum_{i=1}^{m}  \sum_{s=1}^{d_i} \sum_{t=1}^{d_i} \lambda_i \ket{s_A^i t_B^i} \delta_{st} \\ 
		&= \sum_{i=1}^{m}  \sum_{p=1}^{n_i} \lambda_i (\ket{(2p-1)_A^i (2p-1)_B^i} + \ket{(2p)_A^i (2p)_B^i}).
	\end{split}
\end{equation}
If we define 
\begin{alignat}{2}
	\ket{0_A^{ip}}&=\ket{(2p-1)_A^i},\qquad & \ket{1_A^{ip}}&=\ket{(2p)_A^i},\\
	\ket{0_B^{ip}}&=\ket{(2p-1)_B^i},\qquad & \ket{1_B^{ip}}&=\ket{(2p)_B^i},
\end{alignat}
then $\ket{\psi}$ takes the form in Eq.\ \eqref{state}.

\section{Concluding remarks}\label{Conclusion}
We have shown that maximal violation of the EBI, by itself, does not certify self-testability; 
additional requirements need to be met.
The extra requirement that Eq.\ \eqref{necessary} should also be satisfied 
makes the experiment self-testing.
That a maximal violation of the EBI does not lead to self-testability is
because transposition of \emph{some} of the components of Alice's 
observables does not affect the statistics but leads to an \emph{inequivalent} experiment.
Similar issues have been pointed out by other 
authors, see, e.g.,\ Refs.\ \cite{McKague2010,Kaniewski2017a}, and it has been suggested that the definition of 
self-testing should be relaxed ``to include this transposition equivalence'' \cite{Kaniewski2017b}.
Then the results in this paper have to be taken into account since in such a relaxation 
we may be losing physically relevant information, as Eq.\ \eqref{stora} shows. 
Alternative approaches to self-testing based on quantification of incompatibility of measurements 
have been proposed \cite{Kaniewski2017a, Chen2016}. 

In addition, we have completely and explicitly characterized the scenarios in which the EBI is maximally violated.
For a pair of qubits, maximal violation requires measurements corresponding to mutually unbiased bases on the  Bloch sphere on one side and to measurements along the diagonals of a dual cube (inscribed in the Bloch sphere) on the other. 
The general case is a superposition of that for the pair of qubits.

In many applications, Bell inequalities are used to guarantee that quantum mechanical systems exhibit desired properties.
The present paper provides information about the EBI which is potentially useful in any situation 
where a maximal violation of the EBI is used as such a resource. 
Examples include a construction for 
device-independent generation of private randomness  proposed by Ac\'{\i}n {\em et al.} \cite{Acin2016}.
We discuss this construction in a companion 
paper \cite{Andersson2017}.

\begin{acknowledgments}
	We thank Mohamed Nawareg and Massimiliano Smania for fruitful discussions,
	J\k{e}drzej Kaniewski and Yeong-Cherng Liang for useful comments on an earlier draft of the paper,
	and Nicolas Gisin for encouraging remarks.
	We also thank P\"{a}r Z.\ Andersson who has produced Fig.\ \ref{Figure}. 
	AC acknowledges support from Project No.\ FIS2014-60843-P, ``Advanced Quantum Information'' (MINECO, Spain), with FEDER funds,
	the FQXi Large Grant ``The Observer Observed: A Bayesian Route to the Reconstruction of Quantum Theory,''
	and the project ``Photonic Quantum Information'' (Knut and Alice Wallenberg Foundation, Sweden).
\end{acknowledgments}


\begin{thebibliography}{10}
	
\bibitem{Bell64}
	J. S. Bell,
	On the Einstein Podolsky Rosen paradox,
	Physics \textbf{1}, 195 (1964).
	
\bibitem{Ekert91}
	A. K. Ekert,
	Quantum Cryptography Based on Bell's Theorem,
	\href{http://dx.doi.org/10.1103/PhysRevLett.67.661}{Phys. Rev. Lett. \textbf{67}, 661 (1991).}
	
\bibitem{BZPZ04}
	\v{C}. Brukner, M. \.{Z}ukowski, J.-W. Pan, and A. Zeilinger,
	Bell's Inequalities and Quantum Communication Complexity,
	\href{http://dx.doi.org/10.1103/PhysRevLett.92.127901}{Phys. Rev. Lett. \textbf{92}, 127901 (2004).}
	
\bibitem{Colbeck06}
	R. Colbeck,
	{\em Quantum and Relativistic Protocols for Secure Multi-Party Computation},
	Ph.D. thesis, University of Cambridge, 2006; 
	\href{http://arxiv.org/abs/0911.3814}{\eprint{arXiv:0911.3814}.}
	
\bibitem{Gisin2009}
	N. Gisin,
	Bell inequalities: Many questions, a few answers,
	in {\em Quantum Reality, Relativistic Causality, and Closing the Epistemic Circle: Essays in Honour of Abner Shimony},
	The Western Ontario Series in Philosophy of Science,
	edited by W. C. Myrvold and J. Christian (Springer, Berlin, 2009), Vol.\ 73, p.\ 125.
	
\bibitem{Acin2016}
	A. Ac\'{\i}n, S. Pironio, T. V\'ertesi, and P. Wittek,
	Optimal randomness certification from one entangled bit,
	\href{http://dx.doi.org/10.1103/PhysRevA.93.040102}{Phys. Rev. A \textbf{93}, 040102(R) (2016).}

\bibitem{Wootters2006}
	W. K. Wootters,
	Quantum measurements and finite geometry,
	\href{http://dx.doi.org/10.1007/s10701-005-9008-x}{Found. Phys. \textbf{36}, 112 (2006).}

\bibitem{McKague2010}
	M. McKague and M. Mosca,
	Generalized self-testing and the security of the 6-state protocol,
	in {\em Theory of Quantum Computation, Communication, and Cryptography},
	Lecture Notes in Computer Science, 
	edited by W. van Dam, V. M. Kendon, and S. Severini (Springer, Berlin, 2010), Vol.\ 6519, p.\ 113.
	
\bibitem{McKague2010thesis}
	M. McKague,
	{\em Quantum Information Processing with Adversarial Devices},
	Ph.D. Thesis, University of Waterloo, 2010;
	\href{https://arxiv.org/abs/1006.2352}{\eprint{arXiv:1006.2352}.}

\bibitem{Andersson2017}
	O. Andersson, P. Badzi\c{a}g, I. Dumitru, and A. Cabello,
	Device-independent certification of two bits of randomness from one entangled bit and the elegant Bell inequality,
	\href{https://arxiv.org/abs/1707.00564}{\eprint{arXiv:1707.00564}.}

\bibitem{Popescu1992}
	S. Popescu and D. Rohrlich,
	Which states violate Bell's inequality maximally?,
	\href{https://doi.org/10.1016/0375-9601(92)90819-8}{Phys. Lett. A \textbf{169}, 411 (1992).}

\bibitem{Mayers1998}
	D. Mayers and A. Yao,
	Quantum cryptography with imperfect apparatus,
	in
	{\em Proceedings of the 39th IEEE Conference on Foundations of Computer Science,  Palo Alto, CA, 1998} (IEEE, New York, 1998).

\bibitem{Mayers2004}
	D. Mayers and A. Yao,
	Self testing quantum apparatus,
	\href{http://www.rintonpress.com/journals/qiconline.html#v4n4}{Quantum Information \& Computation \textbf{4}, 273 (2004)}.

\bibitem{Magniez2006}
	F. Magniez, D. Mayers, M. Mosca, and H. Olliver,
	Self-testing of quantum circuits,
	in {\em Proceedings of ICALP 2006, Part I}, 
	Lecture Notes in Computer Science,
	edited by M. Bugliesi, B. Preneel, V. Sassone, and I. Wegener (Springer, Berlin, 2006), Vol.\ 4051, p.\ 72.

\bibitem{Holevo2011}
	A. Holevo,
	{\em Probabilistic and Statistical Aspects of Quantum Theory},
	(Scuola Normale Superiore Pisa, Pisa, Italy, 2011), p. 55.
	
\bibitem{McKague2012}
	M. McKague, T. H. Yang, and V. Scarani,
	Robust self-testing of the singlet,
	\href{https://doi.org/10.1088/1751-8113/45/45/455304}{J. Phys. A: Math. Theor. \textbf{45}, 455304 (2012).}

\bibitem{Coladangelo2016}
	A. Coladangelo, K. T. Goh, and V. Scarani,
	All pure bipartite entangled states can be self-tested,
	\href{https://doi.org/10.1038/ncomms15485}{Nat. Commun. \textbf{8}, 15485 (2017).}
	
\bibitem{Kaniewski2017a}
	J. Kaniewski,
	Self-testing of binary observables based on commutation,
	\href{https://link.aps.org/doi/10.1103/PhysRevA.95.062323}{Phys. Rev. A \textbf{95}, 062323 (2017).}
	
\bibitem{Kaniewski2017b}
	J. Kaniewski,
	Private communication.

\bibitem{Chen2016}
	S.-L. Chen, C. Budroni, Y.-C. Liang, and Y.-N. Chen,
	Natural Framework for Device-Independent Quantification of Quantum Steerability, Measurement Incompatibility, and Self-Testing,
	\href{https://link.aps.org/doi/10.1103/PhysRevLett.116.240401}{Phys. Rev. Lett. \textbf{116}, 240401 (2016).}
	

\end{thebibliography}
\end{document}